\documentclass[reprint,showpacs,amsfonts,amsmath,amssymb,prb]{revtex4-1}
\usepackage{graphicx}
\usepackage{dcolumn}
\usepackage{amsmath,amstext}
\usepackage{bm}
\usepackage{color}
\usepackage{float}

\begin{document}

\title{ Theoretical investigation of the magnetic dynamics in $\alpha$-RuCl$_3$}

\author{Wei Wang}
\author{Zhao-Yang Dong}
\author{Shun-Li Yu}
\author{Jian-Xin Li}
\email[]{jxli@nju.edu.cn}
\affiliation{National Laboratory of Solid State Microstructures and Department of Physics, Nanjing University, Nanjing 210093, China}
\affiliation{Collaborative Innovation Center of Advanced Microstructures, Nanjing University, China}

\date{\today}

\begin{abstract}
We study the spin-wave excitations in $\alpha$-RuCl$_3$ by the spin-wave theory. Starting from the five-orbital Hubbard model and the perturbation theory, we derive an effective isospin-$1/2$ model in the large Hubbard ($U$) limit. Based on the energy-band structure calculated from the first-principle method, we find that the effective model can be further reduced to the $K$-$\Gamma$ model containing a ferromagnetic nearest-neighbor (NN) Kitaev interaction ($K$) and a NN off-diagonal exchange interaction ($\Gamma$). With the spin-wave theory, we find that the $K$-$\Gamma$ model can give magnetic excitations which is consistent with the recent neutron scattering experiments.
\end{abstract}
\pacs{71.27.+a, 75.10.Jm, 75.25.Dk}
\maketitle

\section{INTRODUCTION}
Currently, considerable attention has been attracted to exotic physics driven by the interplay of the spin-orbital coupling (SOC), crystal field and electronic correlation\cite{Kim2008,Chaloupka2010,Kim2012,Choi2012,Kim2014,Kimchi2014,Becker2015,Chaloupka2016,Sinn2016,WangHu2015,Li2015,Bieri2015}. Especially, in $4d$ or $5d$ transition metal materials, neither the Hubbard interaction $U$ nor the SOC $\lambda$ can solely lead to the insulating behavior. However, the interplay between $U$, $\lambda$  and crystal field $\Delta$ could induce the so-called spin-orbital assisted Mott insulator\cite{Kim2012,Kim2014,Kim2008}. In $d$ orbitals, an electron has total angular momentum $\bm{J}=\bm{s}+\bm{L}$ with orbital angular momentum $\bm{L}$ of five $d$ orbitals and spin angular momentum $\bm{s}$. When $d$ orbitals are subject to an octahedral crystal field circumstance, these states are split into a $t_{2g}$ triplet and an $e_g$ doublet. For the partially filled $d^5$ configuration under large crystal field, the low-energy physics is dominated by the $t_{2g}$ orbitals and it is depicted by a single hole which has an effective orbital angular momentum $l=1$ and an effective total angular momentum $\bm{J}_{\rm eff}=\bm{s}-\bm{L}^\prime$, where $\bm{L}^\prime$ ($\bm{s}$) is the effective orbital (spin) angular momentum of the $t_{2g}$ orbitals. Thus, for a large SOC, the $t_{2g}$ multiplet is divided into a $J_{\rm eff}=3/2$ quartet and a $J_{\rm eff}=1/2$ Kramers doublet with a reduced band width. Therefore, a moderate interaction $U$ can open a Mott gap in the Kramers doublet. The significant consequence of this $J_{\rm eff}=1/2$ Mott insulator state is that its low-energy spin model has been shown to be the Heisenberg-Kitaev (HK) model\cite{Jackeli2009}, in which the celebrated Kitaev interaction is an unusual bond dependent exchange\cite{Kitaev2006}. The pioneer examples are the $5d^5$-iridate compounds A$_2$IrO$_3$(A=Na,Li)\cite{Choi2012,HwanChun2015,Gretarsson2013,Gretarsson2013a,Chaloupka2013,Singh2012,Chaloupka2010,Chaloupka2015,Rau2014} which contain honeycomb lattices with low-spin magnetic ions Ir$^{4+}$ and the edge-sharing octahedral crystal field. Unfortunately, the fact that Ir ions have large neutron absorption cross-section hinders the neutron studies\cite{Choi2012,HwanChun2015}. In addition, the trigonal distortions arouse the controversy about the application of $J_{\rm eff}=1/2$ picture to iridates\cite{Foyevtsova2013}.

Recently, $\alpha$-RuCl$_3$ which is a $4d^5$ analogue of iridates was suggested as another candidate for the realization of the Kitaev interaction term \cite{Majumder2015a,Banerjee2016,Johnson2015,Zhou2016,Pollini1996}. In contrast to iridates, RuCl$_6$ octahedron is much closer to cubic and layers are weakly coupled by van der Waals interactions. Even though the value of SOC is expected to be smaller than that of $5d$ element, the intermediate SOC of Ru$^{3+}$ combined with correlation effects in a narrow Ru$^{3+}$ $d$ band could also lead to the $J_{\rm eff}=1/2$ picture\cite{Plumb2014,Banerjee2016,Kim2016,Koitzsch2016,Sears2015,Majumder2015a}.  Experimentally, due to stacking faults, two different crystalline symmetries have been reported in this compound, including both $P3_112$\cite{Stroganov1957,Plumb2014,Banerjee2016,Kubota2015} ($P3$) and $C2/m$\cite{Brodersen1968,Fletcher1967,Cao2016,Johnson2015} ($C2$) space groups. The  neutron scattering\cite{Johnson2015,Banerjee2016,Sears2015}, X-ray diffraction\cite{Cao2016} and heat capacity\cite{Johnson2015,Banerjee2016,Majumder2015a,Kubota2015} measurements
have pointed towards a zigzag type magnetic order at T$_{N1} \approx 14$ K and T$_{N2} \approx 8$ K which are associated with stacking faults. Moreover, above magnetic ordering temperature the broad continuum scattering is observed not only in inelastic neutron scattering (INS)\cite{Banerjee2016,Banerjee2016b} but also in Raman scattering\cite{Sandilands2015}, which suggests that $\alpha$-RuCl$_3$ may realize Kitaev physics. The INS experiments\cite{Banerjee2016b} suggest that the Kitaev interaction is antiferromagnetic, but below $T_{N1}$ a spin gap near M point is observed\cite{Banerjee2016,Banerjee2016b,Ran2016}, which is not consistent with the theoretical results based on the HK model with an antiferromagnetic Kitaev interaction. Therefore, the HK model is not enough to describe the physics in $\alpha$-RuCl$_3$. Moreover, many theoretical works suggested that the Kitaev interaction is ferromagnetic\cite{Winter2016,Kim2016a,Yadav2016}. Besides, in previous work, the crystal field is expected to be large enough so that one can only take the $t_{2g}$ manifold into account at low energies. However, the crystal field splitting $\Delta$ between $e_{g}$ and $t_{2g}$ orbitals is estimated to be 2.2 eV from the XAS data\cite{Koitzsch2016,Sandilands2016}, which is comparative to Hubbard interaction $U$. Therefore, it is necessary to study the effect of crystal field $\Delta$ on the low energy behavior by including all of the five $d$ orbitals.

In this paper, based on the tight-binding energy bands from the first-principle calculations on $\alpha$-RuCl$_3$, we derive a minimal isospin model which contains only the nearest neighbor (NN) ferromagnetic Kitaev term and isotropic antiferromagentic off-diagonal exchange interaction, by projecting the five-orbital Hubbard model onto the lowest Kramers doublet. By analysing the magnetic interactions, we find that the exchange between $e_g$ and $t_{2g}$ orbitals can enhance the NN ferromagnetic Kitaev interaction $K$ and off-diagonal exchange $\Gamma$, and reduce the NN ferromagnetic Heisenberg interaction $J$. We investigate the magnetic dynamics of this model which is consistent with the results of INS experiments\cite{Banerjee2016,Banerjee2016b,Ran2016} through the SU(2) spin-wave theory\cite{Choi2012,Haraldsen2009}. We further verify the validity of the minimal isospin model through the comparison to the spin-wave excitations calculated from the exchange model containing all of the $J_{\rm eff}=1/2$ and $J_{\rm eff}=3/2$ states by use of the SU(6) spin-wave theory.

The paper is organized as follows. In Sec.~\ref{MEM}, we first introduce the second order perturbation theory and derive an effective exchange model at the strong coupling limit. By analysing the magnetic interactions based on the energy-band structure from the first-principle calculation, we then arrive at a minimal effective exchange model. In Sec.~\ref{SWE}, we introduce the SU(N) spin-wave theory\cite{Dong2016,Muniz2014} and verify the validity of the minimal exchange model by calculating the spin excitation spectrum and the spin-spin correlation functions. Finally, the discussion and summary are given in Sec.~\ref{DAS}.

\section{Minimal effective model }\label{MEM}

\begin{figure}
  \centering
  \includegraphics[width=0.45\textwidth]{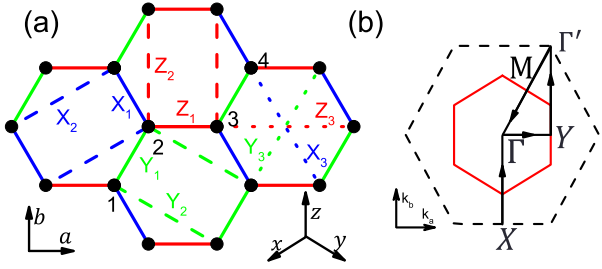}\\
  \caption{(Color online)(a) Lattice structure of Ru$^{3+}$ in $\alpha$-RuCl$_3$. Solid, dashed and dotted lines label first, second and third NN bonds on honeycomb lattice respectively. Red, green and blue colors denote the $Z$, $Y$ and $X$ bonds, respectively. $a, b$ refer to the axes in the honeycomb layer, while $x, y, z$ are the cubic axes of the local octahedron. Sites within a magnetic unit cell for the zigzag order are labeled by $1\sim4$. (b) Structure of the reciprocal space. The red solid lines represent the first Brillouin zone. $\Gamma$, $\Gamma^{\prime}$, $X$, $Y$ and $M$ denote the symmetrical points.}\label{structure}
\end{figure}

We start from the multi-orbital Hubbard model, which includes all of the five $4d$ orbitals of $\mathrm{Ru}^{3+}$ in $\alpha$-RuCl$_3$. It is given as,
\begin{equation}\label{eq1:Hamiltonian}
H=H_{t}+H_{\triangle}+H_{soc}+H_{int}.
\end{equation}
The kinetic energy term $H_{t}$ and crystal field $H_{\triangle}$ are expressed as
\begin{equation}\label{ht}
H_{t}=\sum_{ij,\sigma}\psi^{\dag}_{i\sigma}\mathcal{T}_{ij}\psi_{j\sigma}
\end{equation}
and
\begin{equation}\label{hc}
H_{\triangle}=\sum_{i,\sigma}\psi^{\dag}_{i\sigma}h^{\Delta}_{i}\psi_{i\sigma},
\end{equation}
where $\psi_{i,\sigma}^\dag=( d^\dag_{i,z^2,\sigma}, d^\dag_{i,x^2-y^2,\sigma},d^\dag_{i,yz,\sigma},d^\dag_{i,xz,\sigma},d^\dag_{i,xy,\sigma})$ with $d^\dag_{i,m,\sigma}$ creating an electron of spin $\sigma$ at site $i$ in the orbital $m$. The parameters of $\mathcal{T}_{ij}$ and $h_{i}^{\Delta}$ for a tight-binding fit of the band structure based on the density-functional theory (DFT) are listed in the Appendix~\ref{A}.
$H_{soc}=\sum_i\lambda\bm{L}_i\cdot\bm{s}_i$ is the electron spin-orbital interaction. The on-site Coulomb interaction $H_{int}$ is given by
\begin{align}
H_{int}&=
\frac{1}{2}\sum_{imm'nn'}\sum_{\alpha\beta\mu\nu}\delta_{\alpha\nu}\delta_{\beta\mu}
\{U\delta_{m=m'=n=n'} (1-\delta_{\alpha\beta}) \nonumber \\
&+ U^{\prime}\delta_{mn'}\delta_{m'n}(1-\delta_{mm'}) + J_{H}\delta_{mn}\delta_{m'n'}(1-\delta_{mm'}) \nonumber \\
&+ J^{\prime}\delta_{mm'}\delta_{mn'}(1-\delta_{mn})(1-\delta_{\alpha\beta})\} \nonumber \\
&d^\dag_{im\alpha}d^\dag_{im'\beta}d_{in\mu}d_{in'\nu},
\label{eq:Hu}
\end{align}
where $U$ ($U^{\prime}$) is the intra-orbital (inter-orbital) Coulomb interaction, $J_{H}$ and $J^{\prime}$ are the Hund's coupling and the pairing hopping, respectively. In this paper, we employ $U=U^\prime+2J_{H}$ and $J_{H}=J^\prime$.

Next, we consider the large $U$ limit and derive an effective exchange model through the second-order perturbation approximation. In the perturbation theory, the total Hamiltonian of Eq.~(\ref{eq1:Hamiltonian}) is divided into two parts $H_{0}=H_{int}+H_{soc}+H_\Delta$ and $H_{1}=H_{t}$. Here, $H_{0}$ can be written as $H_{0}=\sum_{i}H_{0i}$ where $H_{0i}$ denotes the Hamiltonian on the site $i$. Then, by projecting out the states in the high-energy subspace with the second-order perturbation approximation, we can obtain the effective Hamiltonian in the low-energy subspace as
\begin{align}\label{eq:Heff}
\mathcal{H}_{\rm eff}=\sum_{ip}E_{ip}|ip\rangle_l {_l}\langle ip| + \sum_{i<j}\mathcal{H}_{ij},
\end{align}
where $E_{ip}$ is the eigenenergy of the $p$-th low-energy eigenstate $|ip\rangle_l$ of $H_{0i}$. Here, the subscript $l$ indicates that the state $|ip\rangle_l$ is in the low-energy subspace of $H_{0i}$. $H_{ij}$ is the effective interaction between the sites $i$ and $j$ by projecting the original Hamiltonian in Eq. (\ref{eq1:Hamiltonian}) into this low-energy subspace, and it can be formally expressed as
\begin{align}
\mathcal{H}_{ij}=&\sum_{pp^\prime p_{1}p_{1}^\prime nn^\prime}\frac{|ip,jp^{\prime}\rangle_l{_l}\langle ip_{1},jp_{1}^\prime|}{2\Delta E_{pp^\prime p_{1}p_{1}^\prime nn'}}\times \nonumber \\
&(\mathcal{H}_{pp',nn'}^{j\rightarrow i}\mathcal{H}_{nn',p_{1}p_{1}'}^{i\rightarrow j}+\mathcal{H}_{pp',nn'}^{i\rightarrow j}\mathcal{H}_{nn',p_{1}p_{1}'}^{j\rightarrow i}),
\label{eqhhh}
\end{align}
\begin{align}
\mathcal{H}_{pp',nn'}^{j\rightarrow i} ={_l}\langle ip,jp'|H_1^{j\rightarrow i}|in,jn'\rangle_h,
\end{align}
\begin{align}
\mathcal{H}_{nn',p_{1}p_{1}'}^{i\rightarrow j} ={_h}\langle in,jn'|H_1^{i\rightarrow j}|ip_{1},jp_{1}'\rangle_l,
\end{align}
\begin{align}
\frac{1}{\Delta E_{pp_1p'p_1^\prime nn'}}=\frac{1}{E_{ip}+E_{jp'}-E_{in}-E_{jn^\prime}}+\nonumber\\
\frac{1}{E_{ip_1}+E_{jp_1'}-E_{in}-E_{jn'}},
\end{align}
where $H_1^{j\rightarrow i}$ is the hopping term from $j$ site to $i$ in $H_{t}$,
$|ip,jp^{\prime}\rangle_l=|ip\rangle_l\otimes|jp^{\prime}\rangle_l$ and $|in,jn'\rangle_h=|in\rangle_h\otimes|jn'\rangle_h$. Here, $|in\rangle_h$ is the $n$-th eigenstate with eigenenergy $E_{in}$ of $H_{0i}$ in the high-energy subspace, and the subscript $h$ indicates that $|in\rangle_h$ is in the high-energy subspace of $H_{0i}$.

In the limit $U\sim\Delta\gg t$ and $\lambda\gg\frac{t^2}{U}$, the local degrees of freedom are governed by the lowest two many-body states of $H_{0i}$, labelled by $\left|1\right\rangle$ and $\left|2\right\rangle$, which become the $J_{\rm eff}=1/2$ Kramers doublet exactly when $\Delta$ tends to infinity and the crystal-field splits in the $t_{2g}$ orbitals (see Appendix \ref{A}) are zero. Thus, we project $H$ into the subspace of the Kramers doublet and expand the Hamiltonian $\mathcal{H}_{\rm eff}$ in the form of $S^\mu_iS_j^\nu(\mu,\nu=0,x,y,z)$, i.e. $\mathcal{H}_{ij}=\sum_{\mu\nu pp'mm'}J_{ij}^{\mu\nu}S_{i,pp'}^\mu S_{j,mm'}^\nu |ip,jm\rangle_l{_l}\langle ip',jm'|$, where $J_{ij}^{\mu\nu}$ is the coefficient of the exchange interaction, $S_i^0$ is the identity matrix, and $S^{\alpha=x,y,z}_{i,pp'}={_l}\langle ip|J_{i,\rm eff}^\alpha|ip'\rangle_l$ is the element of the isospin matrix. The isospin operators satisfy the commutation relation $[S_i^\alpha,S_i^\beta]=i\epsilon^{\alpha\beta\gamma}S_i^\gamma$ ($\epsilon^{\alpha\beta\gamma}$ is Levi-Civita antisymmetry symbol) exactly if $\Delta=\infty$ and the crystal-field splits in the $t_{2g}$ orbitals are zero. Due to the degeneracy of the Kramers doublet, $J_{ij}^{0\alpha}$ and $J_{ij}^{\alpha 0}$ are zeros, and the first term of Eq.~(\ref{eq:Heff}) which is just a constant can be dropped. Therefore, we obtain an effective model involving exchange interactions up to the third NN\cite{supply},
\begin{align}\label{Heff}
H_{\rm eff}&=\sum_{\langle ij\rangle \in\gamma(\alpha\beta)}[J^\gamma\bm{S}_i\cdot\bm{S}_j+K^\gamma S^\gamma_iS^\gamma_j+
  \Gamma^\gamma( S^\alpha_i S_j^\beta \nonumber \\
&+ S^\beta_i S^\alpha_j)+\Gamma^{\prime\gamma}( S^\alpha_i S_j^\gamma+S^\beta_i S^\gamma_j+S^\gamma_i S_j^\beta+S^\gamma_i S^\alpha_j)] \nonumber \\
&+\sum_{\langle\langle ij\rangle\rangle\in\gamma}(J_2^\gamma\bm{S}_i\cdot\bm{S}_j+K_2^\gamma S^\gamma_i S^\gamma_j) \nonumber \\
&+\sum_{\langle\langle\langle ij\rangle\rangle\rangle\in\gamma}K_3^\gamma S^\gamma_i S^\gamma_j.
\end{align}
Here, $\langle ij\rangle$, $\langle\langle ij\rangle\rangle$ and $\langle\langle\langle ij\rangle\rangle\rangle$ denote the first NN, second NN and third NN bonds respectively. $\gamma$ represents the direction of each bond as shown in Fig.~\ref{structure}. For $Z$-type, $X$-type and $Y$-type bonds in Fig.~\ref{structure}, $(\alpha\beta)$'s are $(xy)$, $(yz)$ and $(zx)$ respectively. $J$ and $K$ are the magnitude of the Heisenberg and Kitaev interactions, $\Gamma$ and $\Gamma^{\prime}$ are the off-diagonal exchanges. The second and third NN exchange interactions are generally smaller than the first NN interactions since their hopping integrals are much smaller than the first NN ones (see Appendix \ref{B}), so only the main terms of the second and third NN exchange interactions are retained in Eq. (\ref{Heff}).
In the case of the $P3$ space group, the interactions are invariant for different directions due to the $C_3$ symmetry. However, for the low symmetric $C2$ space group, the interactions on the $X$ and $Y$-bonds are equal but different from those on the $Z$-bonds. Moreover, in the case of the $C2$ space group, to make the $J$ and $\Gamma^{\prime}$ terms on the $X$-type and $Y$-type bonds equal for the spin directions, we will take their average values\cite{Winter2016}.

\begin{figure}
  \centering
  \includegraphics[clip,width=0.45\textwidth]{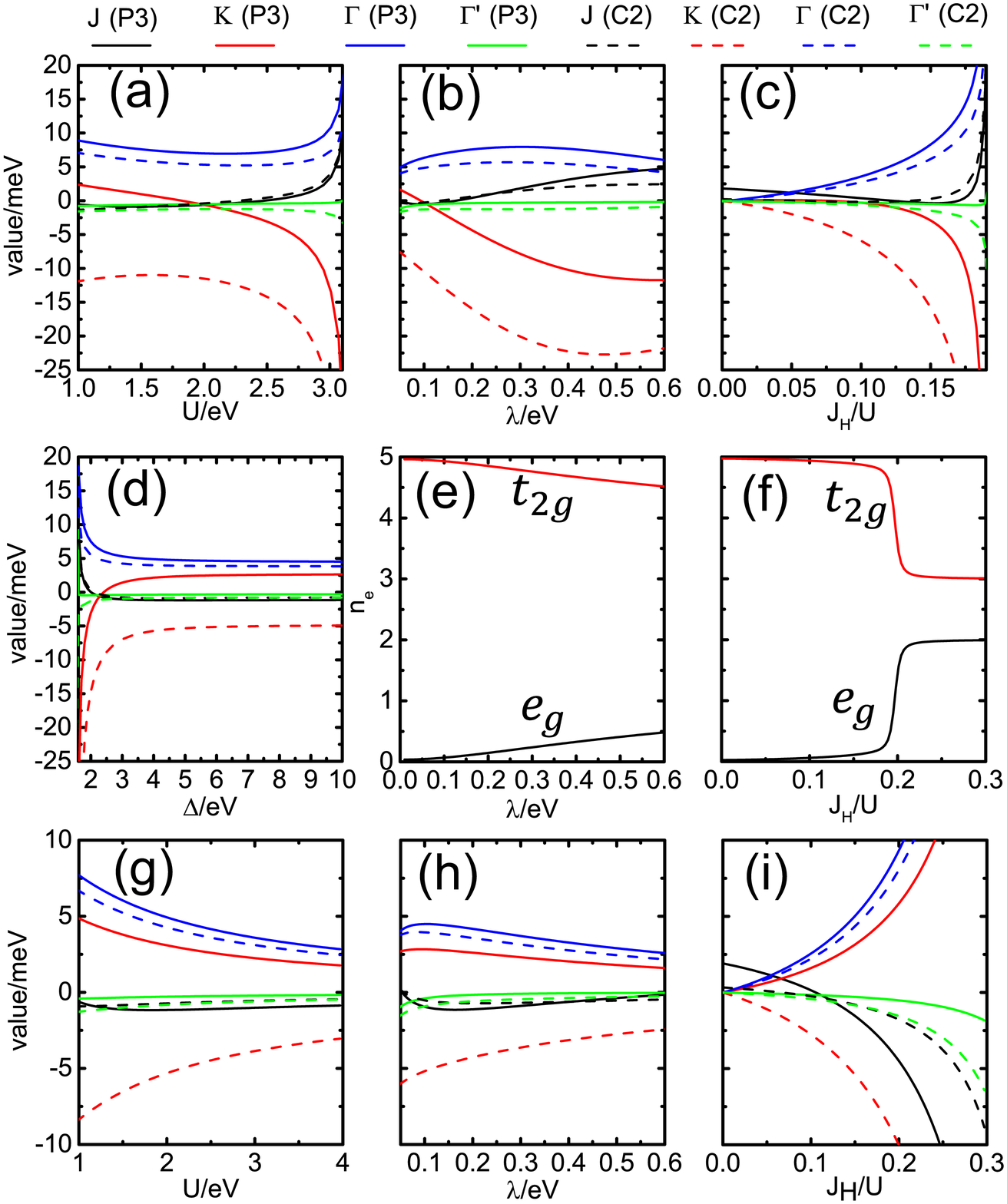}\\
  \caption{(Color online)  Dependence of the first NN interactions on the parameters for $C2$ case (dashed) and $P3$ case (solid). (a) $U$-dependence with $\Delta=2.10$ eV, $\lambda=0.14$ eV and $\frac{J_H}{U}=0.14$. (b) $\lambda$-dependence with $\Delta=2.10$ eV, $U=2.31$ eV and $\frac{J_H}{U}=0.14$. (c) $\frac{J_H}{U}$-dependence with $\Delta=2.10$ eV, $U=2.31$ eV and $\lambda=0.14$ eV. (d) $\Delta$-dependence  with $U=2.31$ eV, $\lambda=0.14$ eV and $\frac{J_H}{U}=0.14$. (g)-(i) with $\Delta=210$ eV corresponding to (a)-(c). (e) and (f) show the number of electrons in Kramers doublet per site corresponding to (b) and (c) respectively. The black (red) line is $e_g$ ($t_{2g}$) orbitals. }\label{JKG1}
\end{figure}
The exchange interaction parameters in Eq.~(\ref{Heff}) depend on the hopping integrals between various orbitals, crystal field $\Delta$, SOC $\lambda$, Hubbard interaction $U$ and Hund's coupling $J_H$. The hopping integrals are determined from the first-principle calculations as listed in Appendix \ref{A}. The dependences of the exchange interactions on $\Delta$, $\lambda$, $U$ and $J_H$ are shown in Fig.~\ref{JKG1}. To simplify the comparison, the values of interactions are bond-averaged in the $C2$ case, so the superscript $\gamma$ is omitted. As the second and third NN terms are small in contrast to the first NN terms, we only present the values of the first NN terms in Fig.~\ref{JKG1}. In Fig.~\ref{JKG1} (a)-(c), we fix $\Delta=2.1$ eV which is suitable for $\alpha$-RuCl$_3$. Their $\Delta$-dependences are then presented in Fig.~\ref{JKG1} (d). To investigate the effect of the $e_g$ orbital, we deliberate to choose an unrealistic large $\Delta=210$ eV and the results are shown in Fig.~\ref{JKG1} (g)-(i) for a comparison.

The noticeable overall feature in Fig.~\ref{JKG1} (a)-(d) is that the Heisenberg exchange term is much smaller than other terms in an extended range of parameters for $\lambda<0.15~\mathrm{eV}$ which is the estimated maximum value for $\lambda$.\cite{Koitzsch2016,Sandilands2016,Banerjee2016,Winter2016,Kim2015} This arises from the nearly offset between the contributions to the $J$-term from the inter-band $e_{g}$-$t_{2g}$ superexchange channels and intra-band $t_{2g}$ channels.\cite{Chaloupka2013} Fig.~\ref{JKG1} (a) shows that the magnitude of the exchange interactions has a trend to decrease and then increase with the increase of $U$. As $U$ increases the gap between the Kramers doublet and other excited states, the effective exchange interactions will decrease with $U$ according to Eq. (\ref{eqhhh}). In Fig.~\ref{JKG1} (a), we fix the value of $J_{H}/U$, so the Hund's coupling $J_{H}$ increases with $U$. For the $4d^{5}$ electron configuration, the Hund's coupling will decrease the energies of the excited states which contain a large weight of $e_{g}$ orbitals, so the exchange interactions increase with $J_{H}$. Therefore, there is a competing relation between $U$ and $J_{H}$ in determining the exchange interactions. We can see this point more clearly in Fig.~\ref{JKG1} (g), where the crystal filed $\Delta$ is set at a deliberate large value, so that the effect of the $e_{g}$ orbitals is excluded and the effect of $J_{H}$ is suppressed. In this case, all the exchange interactions decrease with $U$. In Fig.~\ref{JKG1} (i), the large $J_H/U$ induces the ferromagnetic $J$ interaction and enhances the values of the antiferromagnetic $K$ interaction in the $P3$ case, the ferromagnetic $K$ interaction in the $C2$ case, and the ferromagnetic $\Gamma$ interactions in both cases. The different signs of the $K$ interactions in two cases depend on the hoppings in the $t_{2g}$ orbitals. The antiferromagnetic $K$ term in the $P3$ cases comes mainly from the direct hopping $t_3$ between the $d_{xy}$ orbitals for the $Z$-bond. The $K$ term in the $C2$ case is attributed to the indirect hopping $t_2$ between $t_{2g}$ orbitals via chlorine ions. However, when $\Delta$ is reduced to be comparable to the Hubbard $U$, as shown in panel (c), a large $J_H$ leads to the ferromagnetic $K$ in the $P3$ case and the antiferromagnetic $J$ in both cases. It is also supported by their $\Delta$ dependence.  This is because a large $J_H/\Delta$ increases the mixing of the $e_g$ and $t_{2g}$ orbitals in the Kramers doublet, as shown in panels  (f) where the number of electrons $n_e$ in the $e_g$ orbitals increases rapidly for $J_H/U>0.19$. Moreover, by comparing Fig.~\ref{JKG1} (c) with (i), we find that the exchange channels between the $e_g$ and $t_{2g}$ orbitals can enhance the magnitude of the $\Gamma$ and $K$ interactions. When the weight of $e_g$ orbitals in the Kramers doublet increases rapidly, the values of interactions are divergent and the $J_{\rm eff}=1/2$ picture is no longer applicable. From the $\lambda$ dependence as shown in Fig.~\ref{JKG1} (b), (e) and (h), we can see that in the large $\Delta$ limit the values of interactions are suppressed with $\lambda$ owning to the enhancement of the gap between the $ J_{\rm eff}=1/2$ and $J_{\rm eff}=3/2$ states, which is consistent with previous work\cite{Winter2016}. If $\Delta$ is reduced, the increase of $\lambda$ results in the same effect as the increase of $J_H/U$, as seen in Fig.~\ref{JKG1} (b) and (e). However, when $\lambda$ is increased to $0.3$ eV, the values of interactions increase slowly and even decrease. This is because the effect of the gap between the $J_{\rm eff}=1/2$ and $J_{\rm eff}=3/2$ states on interactions is greater than that of the $e_g$-$t_{2g}$ channels.

In $\alpha$-RuCl$_3$, $\Delta\approx2.2~\mathrm{eV}$, $U=2\sim3~\mathrm{eV}$ and $\lambda=0.13\sim0.15~\mathrm{eV}$ \cite{Koitzsch2016,Sandilands2016,Banerjee2016,Winter2016,Kim2015}, so we find that the leading exchange interactions are $K$ and $\Gamma$ terms according to what we discuss above. Thus, we arrive at a minimal exchange model,
\begin{equation}\label{eq:Hmin}
\begin{split}
  H_{min}=&\sum_{\langle ij\rangle \in\gamma(\alpha\beta)}[K^{\gamma}S^\gamma_i S^\gamma_j+
  \Gamma^{\gamma}(S^\alpha_i S_j^\beta + S^\beta_i S^\alpha_j)]
\end{split}
\end{equation}
The symbols are the same as those in Eq.~(\ref{Heff}). In the $P3$ case, the symmetry allows $K^z=K^x=K^y$ and $\Gamma^z=\Gamma^x=\Gamma^y$, while in the $C2$ case $K^x=K^y=K^z+\delta_1$ and $\Gamma^x=\Gamma^y=\Gamma^z+\delta_2$ with a small amount $\delta_1$ and $\delta_2$.

\section{Spin wave excitation }\label{SWE}

\begin{figure}
  \centering
  \includegraphics[width=0.45\textwidth]{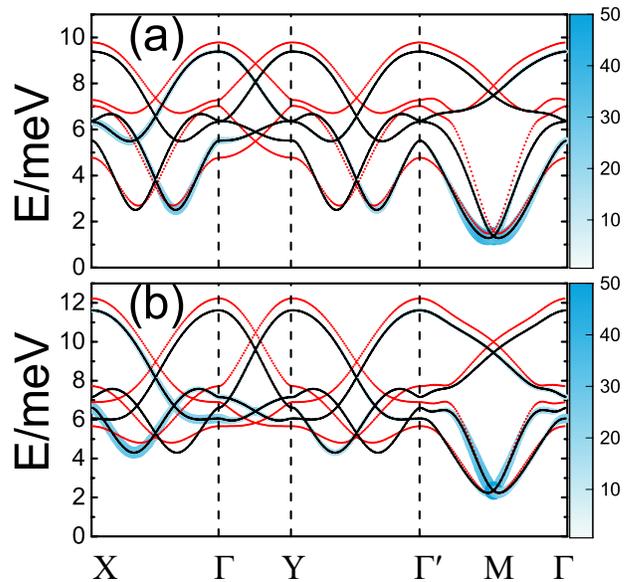}\\
  \caption{(Color online) Spin-wave dispersions along the high symmetry direction $X-\Gamma-Y-\Gamma^\prime-M-\Gamma$ (see Fig.~\ref{structure} (b)) in the $P3$ case (a) and $C2$ case (b). The black solid and red dash lines correspond to the results calculated based on the minimal isospin model in Eq.~(\ref{eq:Hmin}) and the effective isospin model in Eq.~(\ref{Heff}) respectively. The sizes of the colors indicate the magnitude of the isospin correlation function $\tilde{S}$ calculated based on the minimal isospin model in Eq.~(\ref{eq:Hmin}).}\label{sw}
\end{figure}
We now turn to the calculation of the spin-wave excitations. Using the approach of the spin-particle mapping with Schwinger-Wigner bosons\cite{Batista2004}, we map the low-energy state $|ip\rangle_l$ to $b_{ip}^\dagger|0\rangle$ with condition $\sum_{p}b_{ip}^\dagger b_{ip}=1$, where $b_{ip}^\dagger$ creates a boson on site $i$ with quantum number $p$ and $|0\rangle$ is the vacuum state without any bosons. Here, we employ the fundamental irreducible representation of SU($N$) group with $N$ the number of $p$. If the ground state of the system is an ordered state, one of the bosonic modes will condense. Therefore, in the local mean field approximation\cite{Muniz2014}, there exist a stable solution to minimize the ground state energy $\langle G|\mathcal{H}_{eff}|G\rangle$, where $|G\rangle=\prod_{i}\widetilde{b}^\dagger_{i,0}|0\rangle$ is the mean-field ground state represented by the condensed boson $\widetilde{b}^\dagger_{i,0}$ which can be expressed as
\begin{align}\label{eqb}
 \widetilde{b}_{i,0}^\dagger=\sum_pU_{0p}(\bm{x}_i)b_{i,p}^\dagger.
\end{align}
For the case of $J_{\rm eff}=1/2$ discussed above, the local rotation matrix $U_{pp^\prime}(\bm{x}_i)$  depends on two parameters\cite{Choi2012,Haraldsen2009}, i.e. $\bm{x}_i=(\theta_i,\phi_i)$, which are the parameters of the polar coordinates in local frame. For the SU($N$) spin-wave theory, the local rotation matrix has $2(N-1)$ parameters, i.e. $\bm{x}_i=(\theta_{i,1},\cdots,\theta_{i,N-1},\phi_{i,1},\cdots,\phi_{i,N-1})$. When one of the bosons condenses, the corresponding creation and annihilation operators are replaced by\cite{Muniz2014}
\begin{equation}\label{boson}
  \widetilde{b}_{i,0}^\dagger\simeq\widetilde{b}_{i,0}=\sqrt{1-\sum_{p\neq0}\widetilde{b}^\dagger_{i,p}\widetilde{b}_{i,p}}
  =1-\frac{1}{2}\sum_{p\neq0}\widetilde{b}^\dagger_{i,p}\widetilde{b}_{i,p}+\cdots,
\end{equation}
where the $N-1$ bosons $\widetilde{b}_{i,p\neq0}$ become the Holstein-Primakoff bosons now.
By substituting Eq.~(\ref{boson}) and Eq.~(\ref{eqb}) into the Hamiltonian $\mathcal{H}_{\rm eff}$ we obtain the Hamiltonian in terms of rotated bosons as follow,
\begin{equation}\label{Hswt}
  \mathcal{H}_{\rm eff}=\mathcal{H}_0(\{\bm{x}_i\})+\mathcal{H}_1(\{\bm{x}_i\})+\mathcal{H}_2(\{\bm{x}_i\})+\cdots,
\end{equation}
where the subscripts of $\mathcal{H}$ denote the number of rotated bosons.
In the linear spin-wave approximation, we only retain the first three terms of Eq.~(\ref{Hswt}). To find the ground state, we minimize the zero-order term $\mathcal{H}_0(\{\bm{x}_i\})$. When a set of proper parameters $\{\bm{x}_{i}^{0}\}$ are found, the first-order term $\mathcal{H}_1(\{\bm{x}_i\})$ vanishes\cite{Haraldsen2009}. Then, the dispersion is obtained by solving the quadratic term $\mathcal{H}_2(\{\bm{x}_i\})$\cite{supply}.

To search for various possible magnetic ground state including the zigzag order, we choose a magnetic unit cell involving four sites (see Fig.~\ref{structure}) to minimize the ground-state energy. To compare to the INS experiments\cite{Banerjee2016,Banerjee2016b,Ran2016}, we use the SU(N)\cite{Muniz2014} spin-wave theory to calculate the correlation function $\tilde{S}(\bm{q},\omega)$(zero temperature), which is defined as
 \begin{equation}\label{eq:Q}
   \tilde{S}(\bm{q},\omega)=\frac{1}{N}\sum_{ij}e^{\mathrm{i}\bm{q}(\bm{r_i}-\bm{r_j})}\int_{-\infty}^{\infty}\left\langle\bm{Q}_{i}\bm{Q}_{j}(t)\right\rangle e^{-i\omega t}dt,
 \end{equation}
with $\bm{Q}_j(t)=e^{{i}Ht}\bm{Q}_je^{-{i}Ht}$. For the effective and minimum isospin models, the correlation function of the isospin operator $\bm{Q}_{i}=\sum_{pp'=1,2}{_l}\langle p|J_{i,\rm{eff}}|p'\rangle_l b^\dagger_p b_{p'}$ is calculated by the SU(2) spin-wave theory.

\begin{figure}
  \centering
  \includegraphics[width=0.45\textwidth]{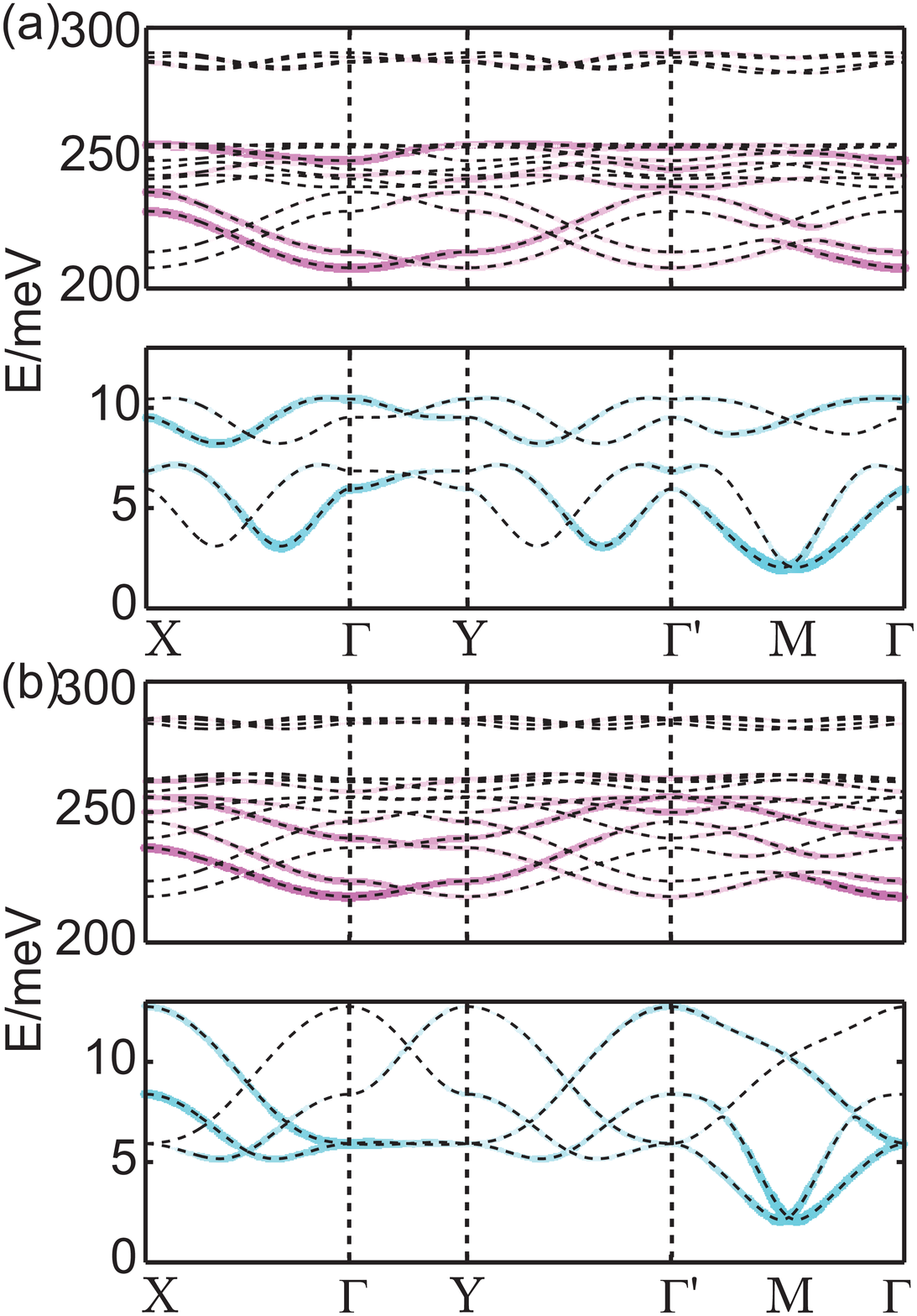}\\
  \caption{(Color online) Spin-wave dispersion (dashed) for the complex effective exchange model in (a) and (b) corresponding to the $P3$ case and $C2$ case, respectively. The sizes of the colors indicate the magnitude of correlation function $\tilde{S}$, and the cyan and magenta colors represent the isospin and spin-orbital excitations, respectively. The large gap between $200$ meV and $13$ meV results from SOC. }\label{sixband}
\end{figure}

Firstly, to obtain a suitable values of $K$ and $\Gamma$ in the minimal isospin model in Eq.~(\ref{eq:Hmin}) for $\alpha$-RuCl$_3$, we optimize $K$ and $\Gamma$ to make the low-energy isospin excitations of Eq.~(\ref{eq:Hmin}) to be in accordance with those of Eq.~(\ref{Heff}) obtained through projecting the five-orbital Hubbard model to the Kramers doublet. The hopping parameters in the five-orbital Hubbard model are from the first-principle calculations as listed in Appendix \ref{A}, and the interaction parameters are chosen as $U=2.31$ eV, $J_{H}=0.32$ eV and $\lambda=0.14$ eV, which are appropriate for $\alpha$-RuCl$_3$ \cite{Koitzsch2016,Sandilands2016,Banerjee2016,Winter2016,Kim2015}. Then, we obtain the exchange interaction parameters in Eq.~(\ref{Heff}) which are listed in Appendix \ref{B}. By a comparison of the spin excitation spectrum for the effective exchange model in Eq.~(\ref{Heff}) and the minimal isospin model in  Eq.~(\ref{eq:Hmin}), we find that $K^\gamma=-5.50$ meV and $\Gamma^\gamma=7.60$ meV ($K^z=-10.92$ meV, $K^x=-10.86$ meV, $\Gamma^z=6.20$ meV and $\Gamma^x=6.00$ meV) in Eq.~(\ref{eq:Hmin}) can give a consistent fit to those obtained by Eq.~(\ref{Heff}) in the $P3$ ($C2$) case, as shown in Fig. \ref{sw}. Moreover, the classical ground states of both the minimal isospin model in Eq.~(\ref{eq:Hmin}) and the effective exchange model in Eq.~(\ref{Heff}) show the same zigzag magnetic order for these parameters.

We then perform the calculations of the SU(2) spin-wave theory based on the minimal isospin model to compute the correlation function in Eq.~(\ref{eq:Q}) for the $J_{\rm eff}=1/2$ isospin. The spin-wave Hamiltonian of the minimal isospin model in Eq. (\ref{eq:Hmin}) is listed in Appendix \ref{C}. The results are presented in Fig. \ref{sw}. We find that the isospin excitations show a gap at the $M$ point and the maximal intensity is also near the $M$ point, which agree well with the INS experiments \cite{Banerjee2016,Banerjee2016b,Ran2016}. Moreover, the direction of the magnetic moment $\bm{m}_i=\langle G|\sum_{pp'}{_l}\langle p|\bm{s}_i+\bm{L}_i|p'\rangle_l b^\dagger_{ip} b_{ip'}|G\rangle$, in which $\bm{L}_i$ is the orbital angular moment of the five $d$ orbitals, tilts $36^\circ$ ($48^\circ$) out of the ab plane in $P3$ ($C2$) case, which roughly coincides with the experimental result of Ref. \onlinecite{Cao2016}.

As shown in Fig. \ref{sw} (a) and (b), we can find the dispersions show no qualitative difference in the $P3$ and $C2$ cases, though the maximum intensities of the correlation function $\tilde{S}$ near the $M$ point in the two cases are in different branches. In addition, the gaps of the isospin excitations in the $P3$ and $C2$ cases are also consistent with each other, though the values of $K$ and $\Gamma$ are obviously different.

To further check the validity of the $K$-$\Gamma$ model shown in Eq.~(\ref{eq:Hmin}), we construct a more complex effective exchange model by projecting the five-orbital Hubbard model in Eq.~(\ref{eq1:Hamiltonian}) to the subspace of the lowest six many-body states of the $4d^{5}$ electron configuration. In this enlarged subspace, besides the $J_{\rm eff}=1/2$ doublet, the $J_{\rm eff}=3/2$ quartet is also included. Thus, in addition to the $J_{\rm eff}=1/2$ isospin excitations, there are also the spin-orbital excitations between the $J_{\rm eff}=1/2$ and $J_{\rm eff}=3/2$ states. In this case, we use the SU(6) spin wave theory, in which the local rotation parameters become $\bm{x}_i=(\theta_{i,1},\cdots,\theta_{i,5},\phi_{i,1},\cdots,\phi_{i,5})$. We calculate the correlation functions in Eq.(\ref{eq:Q}) of the quantities $\bm{Q}_i=\sum_{pp'=1,2}{_l}\langle p|2\bm{s}_{i}+\bm{L}_i|p'\rangle_l b^\dagger_{ip} b_{ip'}$ and $\bm{Q}_i=\sum_{p=1,2}\sum_{p'=3}^6{_l}\langle p|2\bm{s}_{i}+\bm{L}_i|p'\rangle_l b^\dagger_{ip} b_{ip'} +h.c.$, which correspond to the magnetic excitations in the $J_{\rm eff}=1/2$ isospin subspace and those between the $J_{\rm eff}=1/2$ and $J_{\rm eff}=3/2$ states, respectively. Here, $\bm{s}_i$ and $\bm{L}_i$ are the spin and orbital angular momenta, respectively. The factor $2$ of $\bm{s}_i$ is Land\'{e} $g$-factor of spin.
By performing the calculations, we find the ground state of this effective model is of a zigzag spin order, in which the Kramers doublet, $\langle G|\sum_{p=1,2}b_{ip}^\dagger b_{ip}|G\rangle$, has the dominant weight. This result provides a further support to the $J_{\rm eff}=1/2$ isospin picture on which the minimal isospin model is based. More importantly, the low-energy spin-wave excitations (see Fig. \ref{sixband}) calculated from the SU(6) spin-wave theory based on this effective exchange model are also dominated by the $J_{\rm eff}=1/2$ isospin, which is consistent well with those of the minimal isospin model (see Fig. \ref{sw}). Thus, the minimal isospin model in Eq.~(\ref{eq:Hmin}) is suitable for describing the low-energy physics in $\alpha$-RuCl$_3$, and it can be used to investigate other magnetic properties such as the physics of Kitaev spin liquid.  In addition, besides the $J_{\rm eff}=1/2$ isospin excitations, we expect that the spin-orbital excitations between the $J_{\rm eff}=1/2$ and $J_{\rm eff}=3/2$ states revealed by the SU(6) spin-wave calculations in the high-energy parts of Fig.~\ref{sixband} can be observed by the future resonant inelastic X-ray scattering (RIXS) experiments.

\section{DISCUSSION AND SUMMARY}\label{DAS}
\begin{table}
\caption {\label{EXI}Bond-averaged values of magnetic interactions (in meV). $J_3$ represents the third NN Heisenberg interaction. The results from Ref. \onlinecite{Winter2016}, \onlinecite{Kim2016a} and \onlinecite{Kim2015} are also presented for a comparison.}
\centering\def\arraystretch{1.0}
\begin{ruledtabular}
\begin{tabular}{ccccc}
Structure&$J$&$K$&$\Gamma$&$J_3$ \\
\hline
$C2$                  &-0.3        &  -10.9    &  6.1    & 0.03      \\
$P3$                  &0.1          & -5.5    & 7.6    & 0.1     \\
$C2$\cite{Winter2016} &-1.7   & -6.7     & 6.6     & 2.7    \\
$P3$\cite{Winter2016} &-5.5 & 7.6& 8.4 & 2.3 \\
$C2$\cite{Kim2016a}   &-1.0    &-8.2     & 4.2 & - \\
$P3$\cite{Kim2016a,Kim2015}   & -3.5 &4.6  &6.4 &0.8    \\

\end{tabular}
\end{ruledtabular}
\end{table}
We derive a minimal effective isospin model from the five-orbital Hubbard model using the energy bands obtained from the first-principle calculations for $\alpha$-RuCl$_3$. The minimal model contains the ferromagnetic Kitaev term and the antiferromagnetic off-diagonal exchange term. We find that the $e_{g}$-$t_{2g}$ inter-band superexchange channels play an important role in determining the effective exchange interactions on the first NN bonds in $\alpha$-RuCl$_3$.
In the previous works\cite{Winter2016,Kim2016a,Kim2015}, the effects of the $e_g$-$t_{2g}$ mixing on the magnetic interactions have not been investigated in detail. In Ref.~\onlinecite{Winter2016} and Ref.~\onlinecite{Kim2016a}, they only consider the $t_{2g}$ orbitals to study the magnetic interactions and suggest the Kitaev interaction for the $P3$ crystal structure is antiferromagnetic, as shown in Table~\ref{EXI}. Although the authors of Ref.~\onlinecite{Kim2015} suggest that the $e_g$-$t_{2g}$ mixing enhances the antiferromagnetic Kitaev interaction $K>0$ and the ferromagnetic Heisenberg interaction $J<0$, they neglect the intra-atomic exchange interaction between the $e_g$ and $t_{2g}$ orbitals. Here, by considering the Coulomb interactions between all five $d$ orbitals, we find that the $e_g$-$t_{2g}$ mixing induces the ferromagnetic Kitaev coupling $K<0$ in both the $C
2$ and $P3$ crystal structures and reduces the Heisenberg interaction $J$ in both structures. Compared with the previous studies, the third NN Heisenberg interaction $J_3$ is also largely suppressed. This is caused by the different signs of the third NN diagonal hopping integrals in the $t_{2g}$ orbitals as discussed in Appendix~\ref{B}. If the signs are all minus, the third NN Heisenberg interaction $J_3$ is greater than the third NN Kitaev interaction $K_3$, as shown in Appendix~\ref{B}, which is consistent with the result from Ref.~\onlinecite{Winter2016}.

Based on this effective isospin model we investigate the spin-wave excitation using the linear spin-wave theory and find it is consistent with the recent neutron scattering on $\alpha$-RuCl$_3$\cite{Ran2016}, especially the gap opening in the magnon dispersion. In our minimal $K$-$\Gamma$ model, the basic reason of the gap opening is that it is lack of the continuous rotation symmetry, which prevents the Goldstone modes emerging in the magnetic ordering phase. Even though the other perturbing interactions present in the real material, the two exchange interactions in our minimal model still dominates the low-energy physics. Therefore, the gap of the magnon excitation also exists in the real material. Here, the $K$-$\Gamma$ model is a minimal model to describe the magnetic properties of $\alpha$-RuCl$_3$, and it does not completely exclude the possible existence of a small $J$ ( and other terms shown in Fig.~\ref{JKG1}). In fact, from Fig.~\ref{JKG1} we can find that these interaction terms, which are not included in the $K$-$\Gamma$ model, indeed have small non-zero values. However, the magnetic properties are mainly determined by the large $K$ and $\Gamma$ terms, and the results based on the $K$-$\Gamma$ model are consistent with the INS, while the other small terms do not qualitatively affect the results in our spin-wave theory.

\begin{acknowledgments}
This work was supported by the National Natural Science Foundation of China (11374138, 11674158 and 11190023)
and National Key Projects for Research and Development
of China (Grant No. 2016YFA0300401).
\end{acknowledgments}

\appendix
\section{Parameters for tight-binding models and representations of angular momenta}\label{A}

\begin{table}
\caption {\label{TB1}Hopping parameters (in meV) for the first NN. $A$ and $B$ are the sublattice indices, Z$_{1}$ and X$_{1}$ bonds are shown in Fig.~\ref{structure} (a). The results from Ref. \onlinecite{Winter2016}, \onlinecite{Kim2016a} and \onlinecite{Kim2015} are also presented for a comparison.}
\centering\def\arraystretch{1.0}
\begin{ruledtabular}
\begin{tabular}{ccccccc}
Bond&&&\multicolumn{2}{c}{$\mathcal{T}_{ij}$}  \\
   &&$C2$&$P3$&$C2$\cite{Winter2016}&$C2$\cite{Kim2016a}&$P3$\cite{Kim2015} \\
\hline
Z$_1$:                &$d_{z^2}\rightarrow d_{z^2}$        & 22.9     & 23.0     & -       & -      & -  \\
$A$ $\rightarrow$ $B$ &$d_{x^2-y^2}\rightarrow d_{x^2-y^2}$& -30.4    & -91.0    & -       & -      & -      \\
                      &$d_{yz}\rightarrow d_{yz}$          & 42.8     & 61.4     & 50.9    & 36.0   & 65.0 \\
                      &$d_{xz}\rightarrow d_{xz}$          & 42.8     & 61.4     & 50.9    & 36.0   & 66.0  \\
                      &$d_{xy}\rightarrow d_{xy}$          & -117.1   & -206.9   & -154.0  & -62.0 & -229.0   \\
                      &$d_{z^2}\leftrightarrow d_{x^2-y^2}$& 0.0      & 0.0      & -       & -      & -   \\
                      &$d_{z^2}\leftrightarrow d_{yz}$     & 13.8     & -0.7     & -       & -      & -   \\
                      &$d_{z^2}\leftrightarrow d_{xz}$     & 13.8     & -0.7     & -       & -      & -   \\
                      &$d_{z^2}\leftrightarrow d_{xy}$     & 268.0    & 212.4    & -       & -      & -   \\
                      &$d_{x^2-y^2}\leftrightarrow d_{yz}$ & -7.5     & -0.3     & -       & -      & -   \\
                      &$d_{x^2-y^2}\leftrightarrow d_{xz}$ & 7.5      & 0.3      & -       & -      & -   \\
                      &$d_{x^2-y^2}\leftrightarrow d_{xy}$ & 0.0      & 0.0      & -       & -      & -   \\
                      &$d_{yz}\leftrightarrow d_{xz}$      & 156.6    & 108.7    & 158.2   & 191.0  & 114.0   \\
                      &$d_{yz}\leftrightarrow d_{xy}$      & -21.4    & -4.5     & -20.2   & -24.0  & -10.0 \\
                      &$d_{xz}\leftrightarrow d_{xy}$      & -21.4    & -4.5     & -20.2   & -24.0  & -10.0   \\
\hline
X$_1$:                &$d_{z^2}\rightarrow d_{z^2}$        & -14.9    & -62.5     & -       & -      & -  \\
$A$ $\rightarrow$ $B$ &$d_{x^2-y^2}\rightarrow d_{x^2-y^2}$& 11.4     & -11.3     & -       & -      & -      \\
                      &$d_{yz}\rightarrow d_{yz}$          & -104.5   & -206.9    & -103.1  & -75.0 & -229.0 \\
                      &$d_{xz}\rightarrow d_{xz}$          & 42.1     & 61.4      & 44.9    & 37.0   & 65.0  \\
                      &$d_{xy}\rightarrow d_{xy}$          & 41.6     & 61.4      & 45.8    & 37.0   & 66.0   \\
                      &$d_{z^2}\leftrightarrow d_{x^2-y^2}$& -16.8    & -49.4     & -       & -      & -   \\
                      &$d_{z^2}\leftrightarrow d_{yz}$     & -137.3   & -106.2    & -       & -      & -   \\
                      &$d_{z^2}\leftrightarrow d_{xz}$     & 2.3      & -0.6      & -       & -      & -   \\
                      &$d_{z^2}\leftrightarrow d_{xy}$     & -9.8     & 0.1       & -       & -      & -   \\
                      &$d_{x^2-y^2}\leftrightarrow d_{yz}$ & 232.7    & 183.9     & -       & -      & -   \\
                      &$d_{x^2-y^2}\leftrightarrow d_{xz}$ & 11.7     & -0.5      & -       & -      & -   \\
                      &$d_{x^2-y^2}\leftrightarrow d_{xy}$ & 2.8      & -0.8      & -       & -      & -   \\
                      &$d_{yz}\leftrightarrow d_{xz}$      & -12.7    & -4.5      & -15.1   & -26.0  & -10.0   \\
                      &$d_{yz}\leftrightarrow d_{xy}$      & -13.2    & -4.5      & -10.9   & -26.0  & -10.0 \\
                      &$d_{xz}\leftrightarrow d_{xy}$      & 159.9    & 108.7     & 162.2   & 182.0  & 114.0   \\
\end{tabular}
\end{ruledtabular}
\end{table}
The electronic structure calculations are performed with the generalized gradient approximation for the exchange-correlation functional as implemented in Quantum ESPRESSO package\cite{Giannozzi2009} based on the density-functional theory (DFT). To avoid double counting of the SOC\cite{Winter2016}, the SOC were not included in these calculations.
The five-orbital parameters (TB5) in the hopping matrix $\mathcal{T}_{ij}$ from the maximally-localized Wannier orbital\cite{Marzari1997} calculation are shown in Table~\ref{TB1}, Table~\ref{TB2} and Table~\ref{TB3} for the first, second and third NN, respectively. Here, only the parameters along the $Z,X$-type bonds are shown, for other bonds in the $P3$ ($C2$) case, the holistic hopping matrix can be recovered by applying inversion operations and $C_3$ rotations along the $c$-axis perpendicular to the ab plane ($C_2$ rotations along the $Z_1$-bond).
For comparison, we also list the values of hopping integrals from several previous works\cite{Winter2016,Kim2016a,Kim2015}, which only have three-orbital parameters.
In our $C2$ case, the crystal structure is from Ref.~\onlinecite{Cao2016}. In our $P3$ case, an ideal chlorine octahedron is considered and the lattice constants are fixed at $a_0=5.97$ {\AA}, $b_0=5.97$ {\AA} and $c_0=17.2$ {\AA}\cite{Banerjee2016}. The electron operators are expressed as $\psi_{i,\sigma}^\dag=( d^\dag_{i,z^2,\sigma}, d^\dag_{i,x^2-y^2,\sigma},d^\dag_{i,yz,\sigma},d^\dag_{i,xz,\sigma},d^\dag_{i,xy,\sigma})$ and $\psi_{i,\sigma}^\dag=(d^\dag_{i,yz,\sigma},d^\dag_{i,xz,\sigma},d^\dag_{i,xy,\sigma})$ for five-orbital and three-orbital models, repsectively.
 The crystal field in the $P3$ case is given by
\begin{equation}\label{eq:hd}
  h^\Delta_{i}=\left(
             \begin{array}{ccccc}
               \Delta & 0 & \Delta_2^\prime & \Delta_2^\prime & -2\Delta_2^\prime \\
               0 & \Delta & -\sqrt{3}\Delta_2^\prime & \sqrt{3}\Delta_2^\prime & 0 \\
               \Delta_2^\prime & -\sqrt{3}\Delta_2^\prime & 0 & \Delta_3^\prime & \Delta_3^\prime \\
               \Delta_2^\prime & \sqrt{3}\Delta_2^\prime & \Delta_3^\prime & 0 & \Delta_3^\prime \\
               -2\Delta_2^\prime & 0 & \Delta_3^\prime & \Delta_3^\prime & 0 \\
             \end{array}
           \right)
\end{equation}
with $\Delta=1980$ meV, $\Delta_2^\prime=15$ meV and $\Delta_3^\prime=-8.6$ meV. The matrix representation is the same as that defined in Eq.~\ref{hc}. Due to the high symmetry in the $P3$ case, there is only one kind of the crystal field split in the $t_{2g}$ orbitals. However, the low symmetry in the $C2$ case allows three kinds of the crystal field splits in the $t_{2g}$ orbitals, as shown in the next text.
For the $C2$ case, the crystal field is written as
\begin{equation}\label{eq:hd1}
  h^\Delta_{i}=\left(
             \begin{array}{ccccc}
               \Delta+4.4 & 0 & 8.1 & 8.1 & -64.2 \\
               0 & \Delta & -60.9 & 60.9 & 0 \\
               8.1 & -60.9 & 0 & \Delta_1 & \Delta_2 \\
               8.1 & 60.9 & \Delta_1 & 0 & \Delta_2 \\
               -64.2 & 0 & \Delta_2 & \Delta_2 & \Delta_3 \\
             \end{array}
           \right)
\end{equation}
with $\Delta=2272.5$ meV, $\Delta_1=-8.1$ meV, $\Delta_2=-7.0$ meV and $\Delta_3=-3.4$ meV.
The orbital angular momenta in the five-orbital model are expressed as
 \begin{align}
   L^x = &\left(
         \begin{array}{ccccc}
           0 & 0 & {i}\sqrt{3} & 0 & 0 \\
           0 & 0 & {i} & 0 & 0 \\
           -{i}\sqrt{3} & -{i} & 0 & 0 & 0 \\
           0 & 0 & 0 & 0 & {i} \\
           0 & 0 & 0 & -{i} & 0 \\
         \end{array}
       \right),
 \end{align}
  \begin{align}
   L^y = &\left(
         \begin{array}{ccccc}
           0 & 0 & 0 & -{i}\sqrt{3} & 0 \\
           0 & 0 & 0 & {i} & 0 \\
           0 & 0 & 0 & 0 & -{i} \\
           {i}\sqrt{3} & -{i} & 0 & 0 & 0 \\
           0 & 0 & {i} & 0 & 0 \\
         \end{array}
         \right),
    \end{align}
  \begin{align}
   L^z = &\left(
          \begin{array}{ccccc}
            0 & 0 & 0 & 0 & 0 \\
            0 & 0 & 0 & 0 & -2{i} \\
            0 & 0 & 0 & {i} & 0 \\
            0 & 0 & -{i} & 0 & 0 \\
            0 & 2{i} & 0 & 0 & 0 \\
          \end{array}
        \right).
 \end{align}
\begin{table}
\caption {\label{TB2}Hopping parameters (in meV) for the second NN. $A$ is the sublattice index, Z$_{2}$ and X$_{2}$ bonds are shown in Fig.~\ref{structure} (a). The results from Ref. \onlinecite{Winter2016}, \onlinecite{Kim2016a} and \onlinecite{Kim2015} are also presented for a comparison.}
\centering\def\arraystretch{1.0}
\begin{ruledtabular}
\begin{tabular}{ccccccc}
Bond&&&\multicolumn{2}{c}{$\mathcal{T}_{ij}$}  \\
   &&$C2$&$P3$&$C2$\cite{Winter2016}&$C2$\cite{Kim2016a}&$P3$\cite{Kim2015} \\
\hline
Z$_2$:                &$d_{z^2}\rightarrow d_{z^2}$        & 12.2     & 4.6      & -       & -      & -  \\
$A$ $\rightarrow$ $A$ &$d_{x^2-y^2}\rightarrow d_{x^2-y^2}$& -1.1     & -2.2     & -       & -      & -      \\
                      &$d_{yz}\rightarrow d_{yz}$          & -5.9     & -0.3     & -4.7    & -      & 0.0 \\
                      &$d_{xz}\rightarrow d_{xz}$          & -5.9     & -0.3     & -4.7    & -      & 0.0  \\
                      &$d_{xy}\rightarrow d_{xy}$          & -4.6     & -0.5     & -0.4    & -      & 0.0   \\
                      &$d_{z^2}\rightarrow d_{x^2-y^2}$    & -5.7     & -1.5     & -       & -      & -   \\
                      &$d_{z^2}\rightarrow d_{yz}$         & -15.5    & -15.9    & -       & -      & -   \\
                      &$d_{z^2}\rightarrow d_{xz}$         & -11.8    & -8.4     & -       & -      & -   \\
                      &$d_{z^2}\rightarrow d_{xy}$         & 73.0     & 65.9     & -       & -      & -   \\

                      &$d_{x^2-y^2}\rightarrow d_{z^2}$& 5.7       & 1.8        & -       & -      & -   \\
                      &$d_{yz}\rightarrow d_{z^2}$     & -11.8     & -8.3       & -       & -      & -   \\
                      &$d_{xz}\rightarrow d_{z^2}$     & -15.5     & -15.8      & -       & -      & -   \\
                      &$d_{xy}\rightarrow d_{z^2}$     & 73.0      & 65.8       & -       & -      & -   \\

                      &$d_{x^2-y^2}\rightarrow d_{yz}$ & -2.7     & -1.3     & -       & -      & -   \\
                      &$d_{x^2-y^2}\rightarrow d_{xz}$ & 4.6      & 3.5      & -       & -      & -   \\
                      &$d_{x^2-y^2}\rightarrow d_{xy}$ & 4.2      & 2.1      & -       & -      & -   \\
                      &$d_{yz}\rightarrow d_{x^2-y^2}$ & -4.6     & -3.6     & -       & -      & -   \\
                      &$d_{xz}\rightarrow d_{x^2-y^2}$ & 2.7      & 1.0      & -       & -      & -   \\
                      &$d_{xy}\rightarrow d_{x^2-y^2}$ & -4.2     & -2.1     & -       & -      & -   \\

                      &$d_{yz}\rightarrow d_{xz}$      & -43.5    & -37.0    & -23.9   & -  & -20.0   \\
                      &$d_{yz}\rightarrow d_{xy}$      & -0.9     & 4.6      & -1.7    & -  & 3.0 \\
                      &$d_{xz}\rightarrow d_{xy}$      & 8.5      & 6.3      & 11.6    & -  & 6.0   \\
                      &$d_{xz}\rightarrow d_{yz}$      & -63.1    & -58.0    & -60.7   & -  & -58.0   \\
                      &$d_{xy}\rightarrow d_{yz}$      & 8.5      & 6.3      & 11.6    & -  & 6.0 \\
                      &$d_{xy}\rightarrow d_{xz}$      & -0.9     & 4.6      & -1.7    & -  & 4.0   \\
\hline
X$_2$:                &$d_{z^2}\rightarrow d_{z^2}$        & 1.9     & -0.4     & -       & -      & -  \\
$A$ $\rightarrow$ $A$ &$d_{x^2-y^2}\rightarrow d_{x^2-y^2}$& 9.9     & 2.8      & -       & -      & -      \\
                      &$d_{yz}\rightarrow d_{yz}$          & -4.3    & -0.5     & -0.4    & -      & 0.0 \\
                      &$d_{xz}\rightarrow d_{xz}$          & -6.7    & -0.3     & -4.5    & -      & 0.0  \\
                      &$d_{xy}\rightarrow d_{xy}$          & -5.5    & -0.3     & -3.2    & -      & 0.0   \\
                      &$d_{z^2}\rightarrow d_{x^2-y^2}$    & -11.5     & -4.6     & -       & -      & -   \\
                      &$d_{z^2}\rightarrow d_{yz}$         & -40.2   & -34.8    & -       & -      & -   \\
                      &$d_{z^2}\rightarrow d_{xz}$         & 9.9     & 9.1      & -       & -      & -   \\
                      &$d_{z^2}\rightarrow d_{xy}$         & 1.4     & 1.2      & -       & -      & -   \\

                      &$d_{x^2-y^2}\rightarrow d_{z^2}$& 0.2       & -1.3      & -       & -      & -   \\
                      &$d_{yz}\rightarrow d_{z^2}$     & -33.1       & -31.1     & -       & -      & -   \\
                      &$d_{xz}\rightarrow d_{z^2}$     & 9.6         & 7.3       & -       & -      & -   \\
                      &$d_{xy}\rightarrow d_{z^2}$     & 3.9         & 7.0       & -       & -      & -   \\

                      &$d_{x^2-y^2}\rightarrow d_{yz}$ & 61.3        & 56.0      & -       & -      & -   \\
                      &$d_{x^2-y^2}\rightarrow d_{xz}$ & -11.3        & -13.1      & -       & -      & -   \\
                      &$d_{x^2-y^2}\rightarrow d_{xy}$ & -13.0       & -9.0     & -       & -      & -   \\
                      &$d_{yz}\rightarrow d_{x^2-y^2}$ & 65.3        & 58.1      & -       & -      & -   \\
                      &$d_{xz}\rightarrow d_{x^2-y^2}$ & -7.4       & -5.4     & -       & -      & -   \\
                      &$d_{xy}\rightarrow d_{x^2-y^2}$ & -13.7       & -14.2      & -       & -      & -   \\

                      &$d_{yz}\rightarrow d_{xz}$      & 10.5     & 6.3    & 11.8     & -  & 6.0   \\
                      &$d_{yz}\rightarrow d_{xy}$      & 0.4      & 4.6   & 1.3      & -  & 4.0 \\
                      &$d_{xz}\rightarrow d_{xy}$      & -44.0    & -37.0  & -24.3    & -  & -20.0   \\
                      &$d_{xz}\rightarrow d_{yz}$      & -0.5     & 4.6    & -1.2     & -  & 3.0   \\
                      &$d_{xy}\rightarrow d_{yz}$      & 8.8      & 6.3    & 8.3      & -  & 6.0 \\
                      &$d_{xy}\rightarrow d_{xz}$      & -63.2    & -58.0  & -59.1    & -  & -58.0   \\
\end{tabular}
\end{ruledtabular}
\end{table}
\begin{table}
\caption {\label{TB3}Hopping parameters (in meV) for the third NN. $A$ and $B$ are the sublattice indices, Z$_{3}$ and X$_{3}$ bonds are shown in Fig.~\ref{structure} (a). The results from Ref. \onlinecite{Winter2016}, \onlinecite{Kim2016a} and \onlinecite{Kim2015} are also presented for a comparison.}
\centering\def\arraystretch{1.0}
\begin{ruledtabular}
\begin{tabular}{ccccccc}
Bond&&&\multicolumn{2}{c}{$\mathcal{T}_{ij}$}  \\
   &&$C2$&$P3$&$C2$\cite{Winter2016}&$C2$\cite{Kim2016a}&$P3$\cite{Kim2015} \\
\hline
Z$_3$:                &$d_{z^2}\rightarrow d_{z^2}$        & -26.1     & -30.6    & -       & -      & -  \\
$A$ $\rightarrow$ $B$ &$d_{x^2-y^2}\rightarrow d_{x^2-y^2}$& 56.9      & 72.8     & -       & -      & -      \\
                      &$d_{yz}\rightarrow d_{yz}$          & 6.6       & 6.4      & -8.2    & -      & -8.0 \\
                      &$d_{xz}\rightarrow d_{xz}$          & 6.6       & 6.4      & -8.2    & -      & -8.0  \\
                      &$d_{xy}\rightarrow d_{xy}$          & -39.9     & -44.2    & -39.5   & -      & -49.0   \\
                      &$d_{z^2}\leftrightarrow d_{x^2-y^2}$& 0.0       & 0.0      & -       & -      & -   \\
                      &$d_{z^2}\leftrightarrow d_{yz}$     & -6.8      & -6.1     & -       & -      & -   \\
                      &$d_{z^2}\leftrightarrow d_{xz}$     & -6.8      & -6.1     & -       & -      & -   \\
                      &$d_{z^2}\leftrightarrow d_{xy}$     & 22.5      & 26.0     & -       & -      & -   \\
                      &$d_{x^2-y^2}\leftrightarrow d_{yz}$ & 4.7       & 5.7      & -       & -      & -   \\
                      &$d_{x^2-y^2}\leftrightarrow d_{xz}$ & -4.7      & -5.7     & -       & -      & -   \\
                      &$d_{x^2-y^2}\leftrightarrow d_{xy}$ & 0.0       & 0.0      & -       & -      & -   \\
                      &$d_{yz}\leftrightarrow d_{xz}$      & -10.6     & -7.5     & -7.4    & -      & -5.0   \\
                      &$d_{yz}\leftrightarrow d_{xy}$      & 12.4      & 9.0      & 11.7    & -      & 9.0 \\
                      &$d_{xz}\leftrightarrow d_{xy}$      & 12.4      & 9.0      & 11.7    & -      & 9.0   \\
\hline
X$_3$:                &$d_{z^2}\rightarrow d_{z^2}$        & 35.3    & 47.0     & -       & -      & -  \\
$A$ $\rightarrow$ $B$ &$d_{x^2-y^2}\rightarrow d_{x^2-y^2}$& -5.1    & -4.8     & -       & -      & -      \\
                      &$d_{yz}\rightarrow d_{yz}$          & -39.9   & -44.2    & -41.4   & -      & -49.0 \\
                      &$d_{xz}\rightarrow d_{xz}$          & 6.3     & 6.4      & -7.9    & -      & -8.0  \\
                      &$d_{xy}\rightarrow d_{xy}$          & 6.4     & 6.4      & -7.5    & -      & -8.0   \\
                      &$d_{z^2}\leftrightarrow d_{x^2-y^2}$& 35.1    & 44.8     & -       & -      & -   \\
                      &$d_{z^2}\leftrightarrow d_{yz}$     & -12.4   & -13.0    & -       & -      & -   \\
                      &$d_{z^2}\leftrightarrow d_{xz}$     & -0.5    & -1.9     & -       & -      & -   \\
                      &$d_{z^2}\leftrightarrow d_{xy}$     & 7.7     & 8.0      & -       & -      & -   \\
                      &$d_{x^2-y^2}\leftrightarrow d_{yz}$ & 18.3    & 22.5     & -       & -      & -   \\
                      &$d_{x^2-y^2}\leftrightarrow d_{xz}$ & -8.6    & -8.1     & -       & -      & -   \\
                      &$d_{x^2-y^2}\leftrightarrow d_{xy}$ & -3.4    & -2.4     & -       & -      & -   \\
                      &$d_{yz}\leftrightarrow d_{xz}$      & 13.1    & 9.0      & 12.7    & -      & 9.0   \\
                      &$d_{yz}\leftrightarrow d_{xy}$      & 12.3    & 9.0      & 10.7    & -      & 9.0 \\
                      &$d_{xz}\leftrightarrow d_{xy}$      & -10.6   & -7.5     & -7.8    & -      & -5.0   \\
\end{tabular}
\end{ruledtabular}
\end{table}

The three-orbital parameters (TB3) for the $P3$ ($C2$) space group are also shown in Table~\ref{table2} (Table~\ref{TBC2}), which are qualitatively consistent with Ref.~\onlinecite{Kim2015} (Ref.~\onlinecite{Winter2016}).
For the three-orbital model in the $P3$ and $C2$ cases, the crystal fields $ h^\Delta_{i}$ are expressed as
\begin{equation}\label{hd3}
  h^\Delta_{i}=\left(
             \begin{array}{ccc}
                0 & \Delta_3^\prime & \Delta_3^\prime \\
                \Delta_3^\prime & 0 & \Delta_3^\prime \\
                \Delta_3^\prime & \Delta_3^\prime & 0 \\
             \end{array}
           \right)
\end{equation}
with $\Delta_3^\prime=-6.6$ meV and
\begin{equation}\label{hdc3}
  h^\Delta_{i}=\left(
             \begin{array}{ccc}
                0 & \Delta_1& \Delta_2\\
                \Delta_1 & 0 & \Delta_2\\
                \Delta_2 & \Delta_2& \Delta_3 \\
             \end{array}
           \right)
\end{equation}
with $\Delta_1=-7.9$ meV, $\Delta_2=-8.4$ meV and $\Delta_3=-3.2$ meV, respectively.
The orbital angular momenta in the three-orbital model are expressed as
\begin{align}
   L^{\prime x} = &\left(
         \begin{array}{ccc}
           0 & 0 & 0 \\
           0 & 0 & i \\
           0 & -i & 0 \\
         \end{array}
       \right), \\
   L^{\prime y} = &\left(
         \begin{array}{ccc}
           0 & 0 & -i \\
           0 & 0 & 0 \\
           i & 0 & 0 \\
         \end{array}
         \right), \\
   L^{\prime z} = &\left(
          \begin{array}{ccc}
            0 & i & 0 \\
            -{i} & 0 & 0 \\
            0 & 0 & 0 \\
          \end{array}
        \right).
 \end{align}

\begin{table}
\caption {\label{table2}Hopping parameters (in meV) for the three-orbital model in the $P3$ case. $A$ and $B$ are the sublattice indices, $Z_1,Z_2$ and $Z_3$ bonds are shown in Fig.~\ref{structure} (a).}
\centering\def\arraystretch{1.0}
\begin{ruledtabular}
\begin{tabular}{ccccc}
Bond&&\multicolumn{2}{c}{$\mathcal{T}_{ij}$}  \\
   &&$d_{yz}$&$d_{xz}$&$d_{xy}$ \\
\hline
Z$_1$:                &$d_{yz}$& 58.7 & 113.9   & -7.0     \\
$A$ $\rightarrow$ $B$&$d_{xz}$& 113.9 & 58.7  & -7.0   \\
                      &$d_{xy}$& -7.0 & -7.0  & -194.1    \\
\hline
Z$_2$:                &$d_{yz}$ & -0.7 & -27.6   & 3.6        \\
$A$ $\rightarrow$ $A$&$d_{xz}$& -51.9 & -0.7   & 6.2        \\
                      &$d_{xy}$& 6.2 & 3.6   & 1.6        \\
\hline
Z$_3$:                &$d_{yz}$ & -6.3 & -4.8   & 10.7      \\
$A$ $\rightarrow$ $B$&$d_{xx}$& -4.8 & -6.3   & 10.7       \\
                      &$d_{xy}$& 10.7 & 10.7   & -43.9       \\
\end{tabular}
\end{ruledtabular}
\end{table}
\begin{table}
\caption {\label{TBC2}Hopping parameters (in meV) for the three-orbital model in the $C2$ case. A and B are sublattice indices, Z$_{1,2,3}$ and X$_{1,2,3}$ bonds are expressed in Fig.~\ref{structure} (a).}
\centering\def\arraystretch{1.0}
\begin{ruledtabular}
\begin{tabular}{ccccc}
Bond&&&\multicolumn{2}{c}{$\mathcal{T}_{ij}$}  \\
   &&$d_{yz}$&$d_{xz}$&$d_{xy}$ \\
\hline
Z$_1$:                  &$d_{yz}$  & 40.7    & 161.9   & -22.9 \\
$A$ $\rightarrow$ $B$    &$d_{xz}$ & 161.9  & 40.7  & -22.9  \\
                      &$d_{xy}$& -22.9  & -22.9  & -101.5   \\
\hline
X$_1$:                 &$d_{yz}$& -90.7    & -19.0   & -17.9     \\
$A$ $\rightarrow$ $B$  &$d_{xz}$& -19.0  & 39.7  & 164.9    \\
                       &$d_{xy}$ & -17.9  & 164.9  & 39.5    \\
\hline
Z$_2$:                &$d_{yz}$& -5.2   & -27.1   & -1.3     \\
$A$ $\rightarrow$ $A$ &$d_{xz}$  & -63.1  & -5.2  & 8.8   \\
                      &$d_{xy}$  & 8.8  & -1.3  & -1.6 \\
\hline
X$_2$:                &$d_{yz}$& -1.8   & 9.4   & -0.7     \\
$A$ $\rightarrow$ $A$ &$d_{xz}$& -1.8  & -5.2  & -26.9   \\
                      &$d_{xy}$ & 8.4  & -62.8  & -4.5 \\
\hline
Z$_3$:                &$d_{yz}$  & -8.4    & -8.5   & 14.2 \\
$A$ $\rightarrow$ $B$ &$d_{xz}$ & -8.5  & -8.4  & 14.2  \\
                      &$d_{xy}$ & 14.2  & 14.2  & -39.5   \\
\hline

X$_3$:               &$d_{yz}$  & -39.6    & 14.5   & 13.8     \\
$A$ $\rightarrow$ $B$ &$d_{xz}$ & 14.5  & -8.6  & -8.3    \\
                      &$d_{xy}$& 13.8  & -8.3  & -8.3    \\
\end{tabular}
\end{ruledtabular}
\end{table}

\begin{figure}
  \centering
  \includegraphics[width=0.48\textwidth]{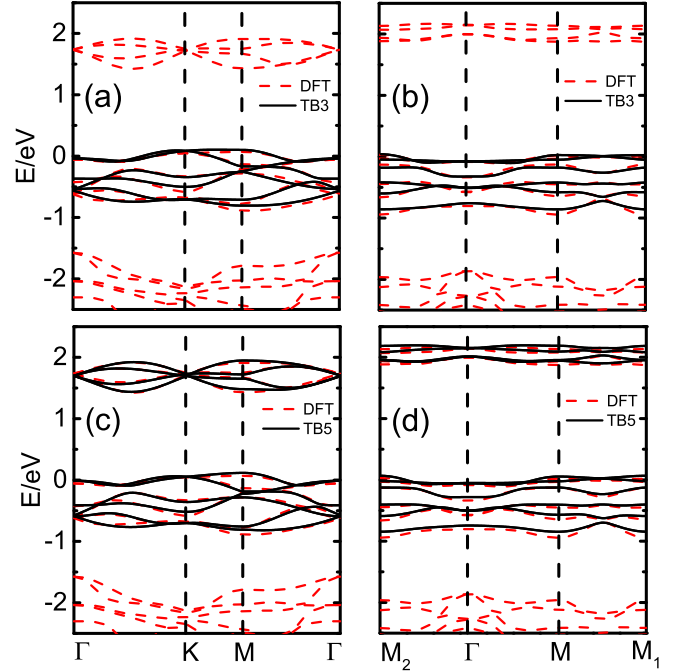}\\
  \caption{(Color online) Band structure of $\alpha$-RuCl$_3$ monolayer. The dashed red lines show the result from DFT without SOC. The tight-binding bands from (a) three-orbital model in the $P3$ case, (b) three-orbital model in the $C2$ case, (c) five-orbital model in the $P3$ case, and (d) five-orbital model in the $C2$ case are denoted by the black solid lines. For the $C2$ case, the high symmetry points $M_1$ and $M_2$ are mid-points of reciprocal lattice vectors. }\label{Afig}
\end{figure}

\begin{figure}
  \centering
  \includegraphics[width=0.48\textwidth]{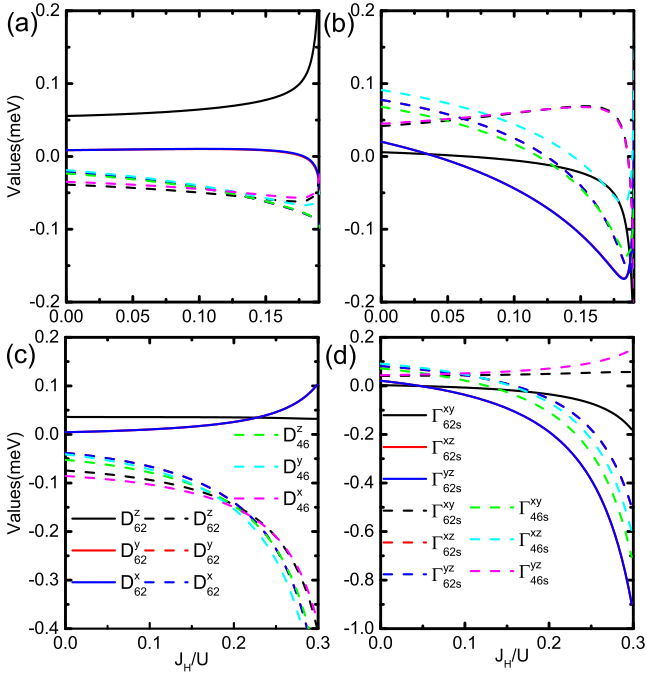}\\
  \caption{(Color online) Dependence of the second NN magnetic interactions on the Hund's coupling $\frac{J_H}{U}$. (a) and (c): the DM interactions; (b) and (d): the symmetrical off-diagonal interactions. $\Delta=2.10$ eV, $U =2.31$ eV and $\lambda = 0.14$ eV are used in (a) and (b). $\Delta=210$ eV, $U =2.31$ eV and $\lambda = 0.14$ eV are used in (c) and (d).  Solid (dash) lines denote the $P3$ ($C2$) case. The subscripts of the magnetic interactions denote the sites, as shown in Fig.~\ref{J3K3} (f).}\label{J2K2}
\end{figure}

\begin{figure}
  \centering
  \includegraphics[width=0.48\textwidth]{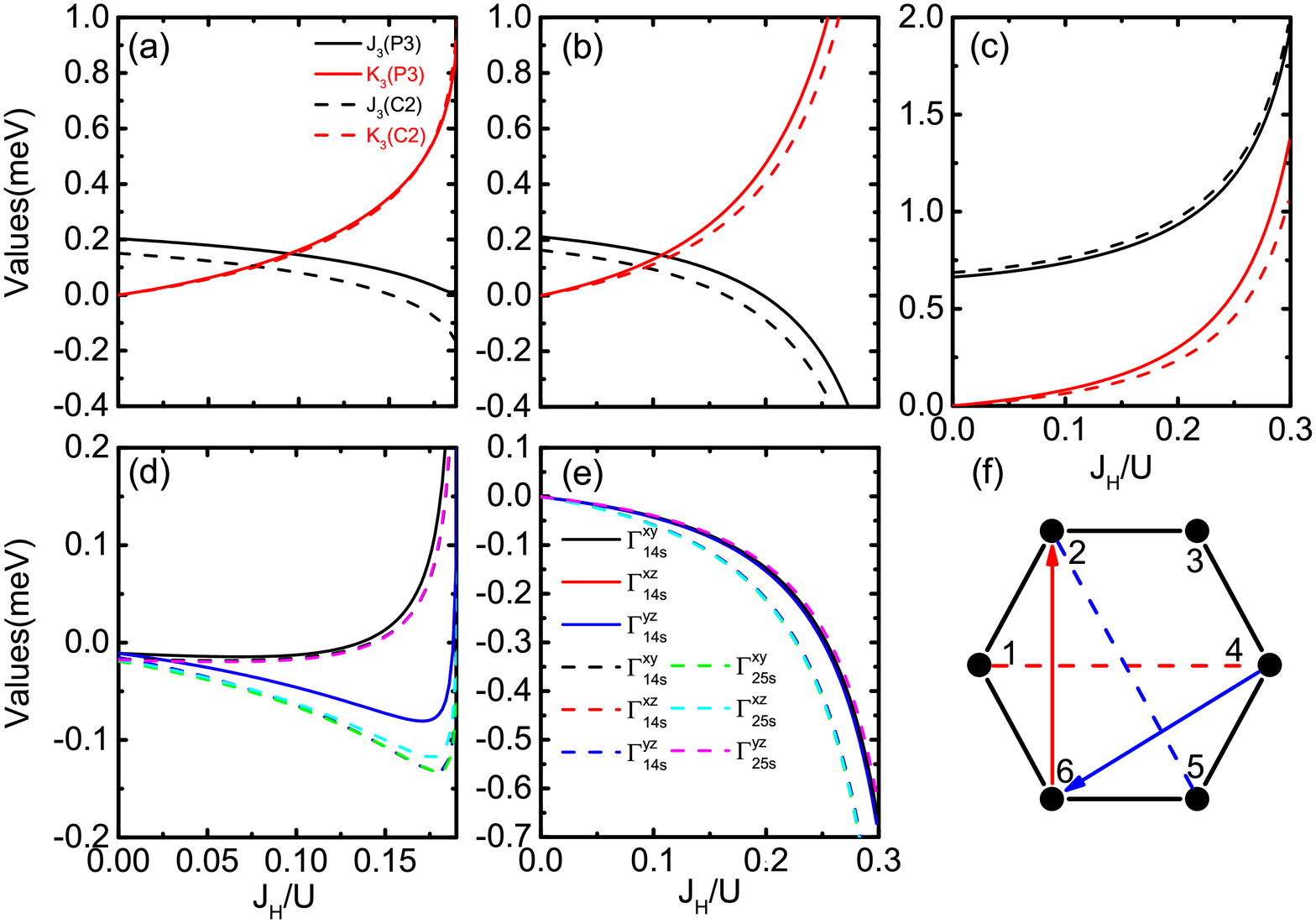}\\
  \caption{(Color online) Dependence of the third NN magnetic interactions on the Hund's coupling $\frac{J_H}{U}$. (a), (b) and (c): the Heisenberg and Kitaev interactions $J_{3}$ and $K_{3}$; (d) and (e): the off-diagonal interactions. $\Delta=2.10$ eV, $U =2.31$ eV and $\lambda = 0.14$ eV are used in (a) and (d). $\Delta=210$ eV, $U =2.31$ eV and $\lambda = 0.14$ eV are used in (b) and (e). $\Delta=\infty$ eV, $U =2.31$ eV and $\lambda = 0.14$ eV are used in (c). Solid (dash) lines denote the $P3$ ($C2$) case. The hopping integrals and crystal fields in the (a), (b), (d) and (e) contain all of the five $d$ orbitals. The hopping parameters in (c) contain only the $t_{2g}$ orbitals. (f): the red and blue solid (dashed) lines represent the second (third) NN $Z$- and $X$- bonds, respectively. The numbers $1$ to $6$ label the sites. }
  \label{J3K3}
\end{figure}
Based on the tight-binding fits, the band structure (black solid) without SOC are shown in Fig.~\ref{Afig}. The red dash lines in Fig.~\ref{Afig} are the band structures from the DFT calculation.

\section{Parameters of the $J_{\rm eff}=1/2$ effective isospin model in Eq.~(\ref{Heff})}\label{B}

The exchange interaction parameters in Eq.~(\ref{Heff}) derived from the five-orbital Hubbard model are calculated based on the tight-binding fit to the DFT calculations and with interactions $U=2.31$ eV, $J_H=0.32$ eV and $\lambda=0.14$ eV. The results are (in meV): $J^\gamma=0.10$, $K^\gamma=-3.35$, $\Gamma^\gamma=7.62$, $\Gamma^{\prime\gamma}=-0.45$, $J_2^\gamma=-0.37$, $K_2^\gamma=0.73$ and $K_3^\gamma=0.42$ ($J^z=-0.40$, $J^{x,y}=-0.23$, $K^{z}=-10.52$, $K^{x,y}=-10.63$, $\Gamma^z=5.07$, $\Gamma^{x,y}=4.75$, $\Gamma^{\prime z}=-1.12$, $\Gamma^{\prime x,y}=-1.14$, $J_2^{z}=-0.31$, $J_2^{x,y}=-0.31$, $K_2^z=0.31$, $K_2^{x,y}=0.31$, $K^z_3=0.31$ and $K^{x,y}_3=0.31$). This set of parameters is used to plot the red dash lines in Fig.~\ref{sw} in the main text.

In Eq.~(\ref{Heff}), we have neglected some terms in the second and third NN exchange interactions, which are found to be much smaller than the NN exchange interactions. Here, we show the $J_{H}$ dependence of the second and third NN exchange interactions neglected in Eq.~(\ref{Heff}) in Fig.~\ref{J2K2} and Fig.~\ref{J3K3}. Because the Hund's coupling $J_{H}$ has the most obvious effect on the exchange interactions as already seen from Fig.~\ref{JKG1}, only the $J_{H}$ dependence is discussed here.

In Fig.~\ref{J2K2}, the Dzyaloshinskii-Moriya (DM) interaction $\bm{D}_{ij}=(D_{ij}^x,D_{ij}^y,D_{ij}^z)$ and off-diagonal $\Gamma$ terms $\Gamma^{\alpha\beta}_{ij,s}$ are shown, which are defined as $D_{ij}^\alpha=(J_{ij}^{\beta\gamma}-J_{ij}^{\beta\gamma})/2$ and $\Gamma_{ij,s}^{\alpha\beta}=(J_{ij}^{\alpha\beta}+J_{ij}^{\beta\alpha})/2$, respectively. The indices $i$ and $j$ are demonstrated in Fig.~\ref{J3K3}.
Comparing Fig.~\ref{J2K2} to Fig.~\ref{JKG1}, we find that the magnitude of the DM interactions and the off-diagonal $\Gamma$ term for the second NN are much smaller than those of the first NN exchange interactions. The reason is that the second NN hopping integrals are much smaller than those for the first NN. Another reason is that the $e_g$-$t_{2g}$ mixing also decreases the DM and $\Gamma$ interactions on second NN bonds.
If we deliberately increase the crystal field to be unrealistic value $\Delta=210$ eV which reduces the mixing of the $e_{g}$ and $t_{2g}$ orbitals,
the magnitudes of the DM and $\Gamma$ exchange interactions are enhanced as shown in Fig.~\ref{J2K2} (c) and (d).

Figure \ref{J3K3} shows the $J_{H}$ dependence of the Heisenberg interactions $J_{3}$, the Kitaev interactions $K_{3}$ and the $\Gamma_{3}$ terms for the third NN bonds. From Fig.~\ref{J3K3} (a) and (b), we find that the effect of the $e_g$-$t_{2g}$ mixing on the third NN diagonal magnetic interactions $J_{3}$ and $K_{3}$ is weak and the Heisenberg interaction $J_3$ is smaller than the Kitaev interaction $K_3$. According to the Eq. (25) in Ref.~\onlinecite{Winter2016}, we have $J_{3}\propto(t_{xz}+t_{yz}+t_{xy})^{2}$, where $t_{xz}$, $t_{yz}$ and $t_{xy}$ are the intra-orbital hopping integrals of the $t_{2g}$ orbitals on the third NN bonds. The signs of these hopping integrals are different (see Table \ref{TB3}), so the intensity of $J_{3}$ is small. However, for the $t_{2g}$ three-orbital model, the signs are the same (see Tables \ref{table2} and \ref{TBC2}), which makes the Heisenberg interaction $J_{3}$ relatively large. Figures~\ref{J3K3} (d) and (e) show that the $e_g$-$t_{2g}$ mixing reduces the third NN off-diagonal $\Gamma$ interactions.

\section{Spin-wave Hamiltonian of the minimal model in Eq.~(\ref{eq:Hmin})}\label{C}

Here, we show the spin-wave Hamiltonian of the minimal model in Eq.~(\ref{eq:Hmin}) by employing the linear spin-wave theory\cite{Choi2012,Haraldsen2009} for the zigzag phase.
In the zigzag phase, we choose the magnetic unit cell $a\times b$ with $a=3\widehat{a}_0$ and $b=\sqrt{3}\widehat{a}_0$, where $\widehat{a}_0$ is the length of the NN bond. For the zigzag order, there is only two degrees of freedom in the magnetic unit cell, i.e. two local rotation parameters $(\theta,\phi)$. Then the zero-order Hamiltonian in the magnetic unit cell is obtained as
\begin{align}
\mathcal{H}_0(\theta,\phi)=\frac{1}{2}(-K^{z}\cos^2(\theta)+\Gamma^{x,y}\sin(2\theta)(\cos(\phi)+\sin(\phi))\nonumber\\
+\sin^2(\theta)(K^{x,y}\sin^2(\phi)-\Gamma^z\sin(2\phi)+K^{x,y}\cos^2(\phi))).
\end{align}
By minimizing the zero-order Hamiltonian, we find the rotation parameter $\phi$ is equal to $\pi/4$ and $\theta$ satisfies $\theta=1/2\tan^{-1}{(-2\sqrt{2}\Gamma^{x,y}/(K^{x,y}+K^z-\Gamma^z))}+\pi/2$. Thus the quadratic Hamiltonian becomes
 \begin{equation}\label{hsw2}
   \mathcal{H}_2=X^\dagger H(\bm{q}) X,
 \end{equation}
where $X^\dagger=(\tilde{b}_{1,\bm{q}}^\dagger,\tilde{b}_{2,\bm{q}}^\dagger,\tilde{b}_{3,\bm{q}}^\dagger,\tilde{b}_{4,\bm{q}}^\dagger,\tilde{b}_{1,-\bm{q}},\tilde{b}_{2,-\bm{q}},\tilde{b}_{3,-\bm{q}},\tilde{b}_{4,-\bm{q}})$ and the number $i$ in the subscript of $\tilde{b}^\dagger_{i,\bm{q}}$ represents the lattice site in the magnetic unit cell (see Fig. \ref{structure}). The matrix $H(\bm{q})$ is given by
\begin{equation}\label{hswq}
  H(\bm{q})=\left(
\begin{array}{cccccccc}
  A & C_1^* & 0 & B & 0 & C^* & 0 & D \\
  C_1^* & A & B^* & 0 & C & 0 & D^* & 0 \\
  0 & B & A & C_1^* & 0 & D_1 & 0 & C^* \\
  B^* & 0 & C_1 & A & D_1^* & 0 & C & 0 \\
  0 & C^* & 0 & D_1 & A & C_1^* & 0 & B \\
  C & 0 & D_1^* & 0 & C_1 & A & B^* & 0 \\
  0 & D & 0 & C^* & 0 & B & A & C_1^* \\
  D^* & 0 & C & 0 & B^* & 0 & C_1 & A
\end{array}
           \right)
\end{equation}
where
\begin{align}
A&=\frac{1}{2}(K^z \cos^2{(\theta)}-\sqrt{2}\Gamma^{x,y}\sin{(2\theta)} \nonumber  \\
&+(\Gamma^z-K^{x,y})\sin^2{(\theta)})\nonumber
\end{align}
\begin{align}
B&=\frac{1}{8}\eta\cos{(\frac{\sqrt{3}q_b}{2})}(3K^{x,y}+K^{x,y}\cos{(2\theta)} \nonumber \\
&-2\sqrt{2}\Gamma^{x,y}\sin{(2\theta)})\nonumber
\end{align}
\begin{align}
C&=\frac{1}{4}\eta^2(-\Gamma^z+K^z)\sin^2{(\theta)}\nonumber\\
C_1&=\frac{1}{4}\eta^2(\Gamma^z+\Gamma^z\cos^2{(\theta)}+K^z\sin^2{(\theta)}) \nonumber
\end{align}
\begin{align}
D&=-\frac{1}{4}\eta(2\sin{(\frac{\sqrt{3}q_b}{2})}(K^{x,y}\cos{(\theta)}+\sqrt{2}\Gamma^{x,y}\sin{(\theta)}) \nonumber\\
&+\cos{(\frac{\sqrt{3}q_b}{2})}\sin{(\theta)}(2\sqrt{2}\Gamma^{x,y}\cos{(\theta)}+K^{x,y}\sin{(\theta)}))\nonumber
\end{align}
\begin{align}
D_1&=\frac{1}{4}\eta(2\sin{(\frac{\sqrt{3}q_b}{2})}(K^{x,y}\cos{(\theta)}+\sqrt{2}\Gamma^{x,y}\sin{(\theta)}) \nonumber \\
&-\cos{(\frac{\sqrt{3}q_b}{2})}\sin{(\theta)}(2\sqrt{2}\Gamma^{x,y}\cos{(\theta)}+K^{x,y}\sin{(\theta)}))\nonumber\\
\eta&=e^{-i\frac{q_a}{2}}
 \end{align}

\bibliography{RuCl3}

\end{document}